\documentstyle[aps, preprint]{revtex}
\def\beq{\begin{eqnarray}}
\def\eed{\end{eqnarray}}
\begin{document}
\draft
\title{Effect of the additional second neighbor hopping on the
charge dynamics in the t-J model}
\author{Feng Yuan$^{1}$ and Shiping Feng$^{1,2,3}$}
\address{$^{1}$Department of Physics, Beijing Normal University,
Beijing 100875, China \\
$^{2}$The Key Laboratory of Beam Technology and Material
Modification of Ministry of Education, Beijing Normal University,
Beijing 100875, China\\
$^{3}$National Laboratory of Superconductivity, Academia Sinica,
Beijing 100080, China}
\maketitle
\begin{abstract}
The effect of the additional second neighbor hopping $t'$ on the
charge dynamics of the $t$-$J$ model in the underdoped regime is
studied within the fermion-spin theory. The conductivity spectrum
of the $t$-$t'$-$J$ model shows the low-energy peak and unusual
midinfrared band, while the resistivity exhibits a nearly
temperature linear dependence with deviation at low temperature
in the underdoped regime. Although the qualitative feature of the
charge dynamics in the $t$-$t'$-$J$ model is the same as in the
case of the $t$-$J$ model, the additional second neighbor hopping
$t'$ leads to a clear shift of the position of the midinfrared
band to the higher energies in the conductivity spectrum, and
suppress the range of the deviation from the temperature linear
dependence in the resistivity.
\end{abstract}
\pacs{71.27.+a, 74.72.-h, 72.10.-d}

It has become clear in the past several years that copper oxide
materials show many unusual normal-state properties \cite{n1}. The
normal-state properties exhibit a number of anomalous properties
in the sense that they do not fit in the conventional Fermi-liquid
theory, and some properties mainly depend on the extent of dopings
\cite{n1}. Among the striking features of the anomalous
normal-state properties stands out the extraordinary charge
dynamics, which is manifested by the optical conductivity and
resistivity \cite{n2}. It has been shown from the experiments
\cite{n1,n2,n15,n16,n17} that the optical conductivity spectrum of
copper oxide materials shows the non-Drude behavior at low energies
and exotic midinfrared band in the charge-transfer gap, while the
resistivity exhibits a linear behavior in the temperature in the
optimally doped regime and a nearly temperature linear dependence
with deviations at low temperatures in the underdoped regime. The
single common feature of copper oxide materials is the
two-dimensional (2D) ${\rm CuO_{2}}$ plane \cite{n1,n2}, and it
seems evident that the anomalous behaviors are governed by this
plane. Since the copper oxide superconductors are doped Mott
insulators, many authors \cite{n3,n4} have suggested that the
essential physics of these materials can be effectively described
by the 2D $t$-$J$ model acting on the space with no doubly occupied
sites, where $t$ is the nearest neighbor hopping matrix element,
and $J$ is the nearest neighbor magnetic exchange interaction. This
model has been used to study the charge dynamics of copper oxide
materials in the underdoped regime, and the results obtained
\cite{n5,n6,n7,n8} from the analytical methods and numerical
simulations are in qualitative agreement with the experiments
\cite{n1,n2,n15,n16,n17}.

However, the recent angle-resolved photoemission spectroscopy
measurements \cite{n9} on copper oxide materials show that
although the highest energy filled electron band is well described
by the $t$-$J$ model in the direction between the $(0,0)$ point and
the $(\pi,\pi)$ point in the momentum space, but both the
experimental data near $(\pi,0)$ point and overall dispersion may
be properly accounted by generalizing the $t$-$J$ model to include
the second- and third-nearest neighbors hopping terms $t'$ and
$t''$. These photoemission results also show that the electron band
width is reduced from the several eV expected from the band theory
to of order $J$, which indicates that the coupling of the electron
to the antiferromagnetic background plays an important role in the
electronic structure \cite{n9}. On the other hand, the charge
response is a powerful probe for the systems of interacting
electrons, and provides very detailed informations of the
excitations, which interact with carriers in the normal-state
\cite{n1,n2,n15,n16,n17}. It is believed that both experiments from
the angle-resolved photoemission spectroscopy measurements and
charge response produce interesting data that introduce important
constraints on the microscopic models and theories. In this case, a
natural question is what is the effect of these additional hoppings
on the charge dynamics of the $t$-$J$ model. In this paper, we
study this issue within $t$-$t'$-$J$ model. Our results indicate
that the conductivity spectrum of the $t$-$t'$-$J$ model shows the
low-energy peak and unusual midinfrared band, while the resistivity
exhibits a nearly temperature linear dependence with deviation at
low temperature in the underdoped regime. Although these
qualitative feature of the charge dynamics in the $t$-$t'$-$J$
model is the same as in the case of the $t$-$J$ model
\cite{n5,n6,n7,n8}, the additional second neighbor hopping $t'$
leads to a clear shift of the position of the midinfrared band to
the higher energies in the conductivity spectrum, and suppress the
range of the deviation from the temperature linear dependence in
the resistivity.

We begin with the $t$-$t'$-$J$ model defined on a square lattice,
\begin{eqnarray}
H=-t\sum_{i\hat{\eta}\sigma}C^{\dagger}_{i\sigma}C_{i+\hat{\eta}
\sigma}+t'\sum_{i\hat{\tau}\sigma}C^{\dagger}_{i\sigma}
C_{i+\hat{\tau}\sigma}+\mu\sum_{i\sigma}C^{\dagger}_{i\sigma}
C_{i\sigma}+J\sum_{i\hat{\eta}}{\bf S}_{i}\cdot
{\bf S}_{i+\hat{\eta}},
\end{eqnarray}
where $\hat{\eta}=\pm \hat{x},\pm\hat{y}$, $\hat{\tau}=\pm \hat{x}
\pm\hat{y}$, $C^{\dagger}_{i\sigma}$ ($C_{i\sigma}$) is the
electron creation (annihilation) operator, ${\bf S}_{i}=
C^{\dagger}_{i}{\bf \sigma}C_{i}/2$ are spin operators with $\sigma
=(\sigma_{x},\sigma_{y}, \sigma_{z})$ as the Pauli matrices, and
$\mu$ is the chemical potential. The strong electron correlation in
the $t$-$J$ and $t$-$t'$-$J$ model manifests itself by this
electron single occupancy on-site local constraint \cite{n3,n4},
and then the on-site local constraint should be treated properly.
Recently a fermion-spin theory based on the charge-spin separation
\cite{n11,n12}, $C_{i\uparrow}=h^{\dagger}_{i}S^{-}_{i}$ and
$C_{i\downarrow}=h^{\dagger}_{i}S^{+}_{i}$, has been proposed to
incorprate this on-site local constraint, where the spinless
fermion operator $h_{i}$ describes the charge (holon) degrees of
freedom, while the pseudospin operator $S_{i}$ describes the spin
(spinon) degrees of freedom. The main advantage of this approach
is that the on-site local constraint can be treated exactly in
analytical calculations. In the fermion-spin representation, the
$t$-$t'$-$J$ model can be expressed \cite{n11,n12} as,
\begin{eqnarray}
H &=& t\sum_{i\hat{\eta}}h^{\dagger}_{i+\hat{\eta}}h_{i}(S^{+}_{i}
S^{-}_{i+\hat{\eta}}+S^{-}_{i}S^{+}_{i+\hat{\eta}}) -
t'\sum_{i\hat{\tau}}h^{\dagger}_{i+\hat{\tau}}h_{i}(S^{+}_{i}
S^{-}_{i+\hat{\tau}}+S^{-}_{i}S^{+}_{i+\hat{\tau}}) \nonumber \\
&-& \mu \sum_{i}h^{\dagger}_{i}h_{i} + J_{eff}\sum_{i\hat{\eta}}
({\bf S}_{i}\cdot {\bf S}_{i+\hat{\eta}}),
\end{eqnarray}
where $J_{eff}=J[(1-\delta)^{2}-\phi^{2}_{1}]$, the holon
particle-hole order parameter $\phi_{1}=\langle h^{\dagger}_{i}
h_{i+\hat{\eta}}\rangle$, and $S^{+}_{i}$ and $S^{-}_{i}$ are the
pseudospin raising and lowering operators, respectively.

Since the on-site electron local constraint has been treated
exactly within the framework of the fermion-spin theory, then the
extra gauge degree of freedom related with the electron on-site
local constraint under the charge-spin separation does not appear
in the fermion-spin theory. In this case, the spin fluctuation
couples only to spinons \cite{n13}, while the charge fluctuation
couples only to holons \cite{n8}, but the strong correlation
between holons and spinons still is considered through the holon's
order parameters entering in the spinon's propagator and the
spinon's order parameters entering in the holon's propagator,
therefore both holons and spinons contribute to the charge and spin
dynamics. Within the fermion-spin theory, the charge dynamics of
the $t$-$J$ model in the underdoped regime has been discussed
\cite{n8} by considering the holon fluctuation around the
mean-field solution, where the holon part is treated by the loop
expansion to the second-order. Following their discussions, we
can obtain the optical conductivity in the present $t$-$t'$-$J$
model as,
\begin{eqnarray}
\sigma (\omega)={1\over 2}(2Ze)^2{1\over N}\sum_k\gamma_{sk}^{2}
\int^{\infty}_{-\infty}{d\omega'\over 2\pi}A_{h}(k,\omega'+\omega)
A_{h}(k,\omega'){n_{F}(\omega'+\omega)-n_{F}(\omega')\over\omega},
\end{eqnarray}
where $Z$ is the number of the nearest neighbor or second-nearest
neighbor sites, $\gamma_{sk}=t\chi_{1}({\rm sin}k_{x}+{\rm sin}
k_{y})/2-t'\chi_{2}({\rm sin}k_{x}{\rm cos}k_{y}+{\rm cos}k_{x}
{\rm sin}k_{y})$, the spinon correlation functions
$\chi_{1}=\langle S_{i}^{+}S_{i+\hat{\eta}}^{-}\rangle$,
$\chi_{2}=\langle S_{i}^{+}S_{i+\hat{\tau}}^{-}\rangle$,
$n_{F}(\omega)$ is the fermion distribution function, and the
holon spectral function $A_{h}(k,\omega)$ is obtained as $A_{h}
(k,\omega)=-2{\rm Im}g(k,\omega)$. The full holon Green's function
$g^{-1}(k,\omega)=g^{(0)-1}(k,\omega)-\Sigma_{h}^{(2)}(k,\omega)$
with the mean-field holon Green's function $g^{(0)-1}(k,\omega)=
\omega-\xi_{k}$ and the second-order holon self-energy from the
spinon pair bubble \cite{n8},
\begin{eqnarray}
\Sigma_{h}^{(2)}(k,\omega)&=&\left({Z\over N}\right)^2\sum_{pp'}
\gamma^{2}_{12} (k,p,p'){B_{p'}B_{p+p'}\over 4\omega_{p'}
\omega_{p+p'}}\times \left ( 2{F_{1}(k,p,p')\over\omega+
\omega_{p+p'}-\omega_{p'}-\xi_{p+k}} \right. \nonumber \\
&+&\left. {F_{2}(k,p,p')\over \omega +\omega_{p'}+\omega_{p+p'}-
\xi_{p+k}}-{F_{3}(k,p,p')\over\omega -\omega_{p+p'}-\omega_{p'}-
\xi_{p+k}}\right ),
\end{eqnarray}
where $\gamma_{12}(k,p,p')=t(\gamma_{{\bf p'-k}}+
\gamma_{{\bf p'+p+k}})-t'(\gamma'_{{\bf p'-k}}+
\gamma'_{{\bf p'+p+k}})$, $\gamma_{{\bf k}}=(1/Z)
\sum_{\hat{\eta}}e^{i{\bf k}\cdot\hat{\eta}}$, $\gamma'_{{\bf k}}
=(1/Z)\sum_{\hat{\tau}}e^{i{\bf k}\cdot\hat{\tau}}$,
$F_{1}(k,p,p')=n_{F}(\xi_{p+k})[n_{B}(\omega_{p'})-n_{B}
(\omega_{p+p'})]+n_{B}(\omega_{p+p'})[1+n_{B}(\omega_{p'})]$,
$F_{2}(k,p,p')=n_{F}(\xi_{p+k})[1+n_{B}(\omega_{p+p'})+n_{B}
(\omega_{p'})]+n_{B}(\omega_{p'})n_{B}(\omega_{p+p'})$,
$F_{3}(k,p,p')=n_{F}(\xi_{p+k)}[1+n_{B}(\omega_{p+p'})+n_{B}
(\omega_{p'})]-[1+n_{B}(\omega_{p'})][1+n_{B}(\omega_{p+p'})]$,
$n_{B}(\omega_{p})$ is the boson distribution functions,
$B_{k}=\Delta_{1}[2\chi^{z}_{1}(\epsilon\gamma_{{\bf k}}-1)+
\chi_{1}(\gamma_{{\bf k}}-\epsilon)]-\Delta_{2}(2\chi^{z}_{2}
\gamma'_{{\bf k}}-\chi_{2})$, $\Delta_{1}=2ZJ_{eff}$, $\Delta_{2}
=4Z\phi_{2}t'$, $\epsilon=1+2t\phi_{1}/J_{eff}$, the mean-field
holon spectrum $\xi_{k}=2Zt\chi_{1}\gamma_{{\bf k}}-2Zt'\chi_{2}
\gamma'_{{\bf k}}-\mu$, and the mean-field spinon spectrum,
\begin{eqnarray}
\omega^{2}_{k}&=&\Delta_{1}^{2}\left([\alpha C^{z}_{1}+{1\over 4Z}
(1-\alpha)-\alpha\epsilon\chi^{z}_{1}\gamma_{k}-{1\over 2Z}\alpha
\epsilon\chi_{1}](1-\epsilon\gamma_{k}) \right. \nonumber \\
&+&\left. {1\over 2}\epsilon[\alpha C_{1}+{1\over 2Z}(1-\alpha)-
\alpha\chi_{1}\gamma_{k}-{1\over 2}\alpha\chi^{z}_{1}](\epsilon-
\gamma_{k})\right ) \nonumber \\
&+&\Delta_{2}^{2}\left ([\alpha\chi^{z}_{2}\gamma'_{k}-{3\over 2Z}
\alpha\chi_{2}]\gamma'_{k}+{1\over 2}[\alpha C_{2}+{1\over 2Z}
(1-\alpha)-{1\over 2}\alpha\chi^{z}_{2}]\right ) \nonumber\\
&+&\Delta_{1}\Delta_{2}\left (\alpha\chi^{z}_{1}\gamma'_{k}
(1-\epsilon\gamma_{k})+{\alpha\over 2}(\chi_{1}\gamma'_{k}-C_{3})
(\epsilon-\gamma_{k})+\alpha\gamma'_{k}(C^{z}_{3}-\epsilon
\chi^{z}_{2}\gamma_{k})\right. \nonumber \\
&-&\left. {1\over 2}\alpha\epsilon(C_{3}-\chi_{2}\gamma_{k})
\right),
\end{eqnarray}
with the spinon correlation functions
$\chi^{z}_{1}=\langle S_{i}^{z}S_{i+\hat{\eta}}^{z}\rangle$,
$\chi^{z}_{2}=\langle S_{i}^{z}S_{i+\hat{\tau}}^{z}\rangle$,
$C_{1}=(1/Z)\sum_{\hat{\eta'}}\langle S_{i+\hat{\eta}}^{+}
S_{i+\hat{\eta'}}^{-}\rangle$,
$C^{z}_{1}=(1/Z)\sum_{\hat{\eta'}}\langle S_{i+\hat{\eta}}^{z}
S_{i+\hat{\eta'}}^{z}\rangle$,
$C_{2}=(1/Z)\sum_{\hat{\tau'}}\langle S_{i+\hat{\tau}}^{+}
S_{i+\hat{\tau'}}^{-}\rangle$,
$C_{3}=(1/Z)\sum_{\hat{\tau}}\langle S_{i+\hat{\eta}}^{+}
S_{i+\hat{\tau}}^{-}\rangle$, $C^{z}_{3}=(1/Z)\sum_{\hat{\tau}}
\langle S_{i+\hat{\eta}}^{z}S_{i+\hat{\tau}}^{z}\rangle$, and the
holon particle-hole order parameter
$\phi_{2}=\langle h^{\dagger}_{i}h_{i+\hat{\tau}}\rangle$.
In order not to violate the sum rule of the correlation function
$\langle S^{+}_{i}S^{-}_{i}\rangle=1/2$ in the case without
antiferromagnetic long-range-order, the important decoupling
parameter $\alpha$ has been introduced in the mean-field
calculation, which can be regarded as the vertex correction
\cite{n12,n14}. All the above mean-field order parameters are
determined by the self-consistent calculation \cite{n12}.

Although the charge dynamics of the copper oxide materials is very
complicated, the transport studies of electron excitations have
revealed much about the nature of the charge carriers
\cite{n1,n2,n15,n16,n17}. In this paper, we are interested in the
effect of the additional second-neighbor hopping on the charge
dynamics of the $t$-$J$ model. To make the discussion simpler, we
only study the charge dynamics of the $t$-$t'$-$J$ model in the
underdoped regime. We have performed the numerical calculation for
the optical conductivity of the $t$-$t'$-$J$ model, and the zero
temperature results with doping concentration (a) $\delta =0.06$
and (b) $\delta=0.12$ for parameter $t/J=2.5$, $t'/J=0.1$ (dashed
line), $t'/J=0.3$ (dotted line), and $t'/J=0.5$ (dash-dotted line)
are plotted in Fig. 1, where we have taken charge $e$ as the unit.
For comparison, the corresponding result \cite{n8} of the $t$-$J$
model (solid line) is also shown in Fig. 1. The conductivity
spectrum of the $t$-$t'$-$J$ model shows the low-energy peak at
$\omega<0.5t$ and midinfrared band appearing inside the
charge-transfer gap of the undoped systems. Although the
qualitative feature of the conductivity spectrum in the
$t$-$t'$-$J$ model is the same as in the case of the $t$-$J$ model
\cite{n5,n8}, the additional second neighbor hopping $t'$ leads to
a clear shift of the position of the midinfrared band to the higher
energies in the conductivity spectrum. Moreover, the
charge-transfer gap in the conductivity is doping dependent,
decreases with increasing dopings. Since the spectral weight from
both the low-energy peak and midinfrared sideband represents the
actual free-carrier density, the suppression of the midinfrared
band with increasing dopings means that only few amount of the
free-carrier are taken from the Drude absorption to the midinfrared
band, which leads to the unusual decay of conductivity at low
energies as $\sigma(\omega)\propto 1/\omega$ in the underdoped
regime, which are consistent with the experiments
\cite{n1,n2,n15,n16,n17}.

The quantity which is closely related to the optical conductivity
is the resistivity, which can be expressed as
$\rho(T)=1/\sigma_{0}(T)$
where the dc conductivity $\sigma_{0}(T)$ can be obtained from Eq.
(3) as $\sigma_{0} (T)=\lim_{\omega\rightarrow 0}\sigma (\omega)$.
This resistivity has been calculated numerically, and the results
at doping concentration $\delta=0.06$ for parameter $t/J=2.5$, (a)
$t'=0$ and (b) $t'/J=0.1$ are shown in Fig. 2(a) and Fig. 2(b),
respectively. From Fig. 2, we find that in the underdoping the
resistivity indeed exhibits a nearly temperature linear dependence
with deviation at low temperature, which also is in agreement with
the experiments \cite{n1,n2,n15,n16,n17}. In the previous study of
the charge dynamics based on the $t$-$J$ model \cite{n8}, it is
shown that the range of the deviation from the temperature linear
dependence in the resistivity in the $t$-$J$ model is suppressed
with increasing dopings. However, our present results based on the
$t$-$t'$-$J$ model show that for the same doping, this range of
the deviation from the temperature linear dependence in the
resistivity is also suppressed with increasing the values of $t'$.

Since the scattering of holons dominates the charge dynamics in
the fermion-spin theory, therefore the present study indicates
that in the underdoped regime the observed crossover from the
temperature linear to the nonlinear range in the resistivity in
low temperatures is closely related to the pseudogap in holon
excitations. This can be understood from the physical property of
the holon density of states (DOS)
$\Omega(\omega)=1/N\sum_{k}A_{h}(k,\omega)$ . The numerical result
of the holon DOS $\Omega(\omega)$ at doping $\delta=0.06$ for
parameters $t/J=2.5$ and $t'=0$ in temperature $T$=0 is shown in
Fig. 3. For comparison, the corresponding mean-field result
(dashed line) is also shown in Fig. 3. From Fig. 3, it is found
that the holon DOS in the mean-field approximation consists of the
central part only, which comes from the noninteracting particles.
After including fluctuations the central part is renormalized and
a V-shape holon pseudogap near the chemical potential $\mu$ in
the underdoped regime appears. For the further understanding the
physical property of this holon pseudogap, we plot the phase
diagram $T^{*}\sim \delta$ at parameters $t/J=2.5$ and $t'=0$ in
Fig. 4, where $T^{*}$ marks the development of the holon pseudogap
in the holon density of states. As seen from Fig. 4, this holon
pseudogap is doping and temperature dependent, and grows
monotonously as the doping $\delta$ decreases, and disappear in
higher doping. Moreover, this holon pseudogap decreases with
increasing temperatures, and vanishes at higher temperatures.
Now we discuss the effect of $t'$ on the pseudogap. In Fig. 5,
we plot the phase diagram $T^{*}\sim t'$ at doping $\delta=0.06$
with parameters $t/J=2.5$, which shows that the holon pseudogap
also decreases with increasing the values of $t'$, and vanishes
for $t'/J>0.1$. The $t$-$J$ model is characterized by a
competition between the kinetic energy ($t$) and magnetic energy
($J$). The magnetic energy $J$ favors the magnetic order for spins,
while the kinetic energy $t$ favors delocalization of holes and
tends to destroy the magnetic order. Our results show that the
additional second neighbor hopping $t'$ in the $t$-$J$ model may
be equivalent to increase the kinetic energy, and its influence on
the charge dynamics of the $t$-$J$ model is similar to the effect
of dopings. This is why at least for small values of $t'$ the
qualitative behavior of the charge dynamics in the $t$-$t'$-$J$
model is the same as these obtained \cite{n5,n6,n7,n8} from the
$t$-$J$ model. Since the full holon Green's function (then the
holon spectral function and DOS) is obtained by considering the
second-order correction due to the spinon pair bubble, therefore
the holon pseudogap is closely related to the spinon fluctuation.
For small dopings and lower temperatures, the holon kinetic energy
is much smaller than the magnetic energy, in this case the
magnetic fluctuation is strong enough to lead to the holon
pseudogap. This holon pseudogap would reduce the holon scattering
and thus is responsible for the deviation from the temperature
linear behavior in the resistivity. With increasing temperatures,
dopings, or values of $t'$ , the holon kinetic energy is
increased, while the spinon magnetic energy is decreased. In the
region where the holon pseudogap closes at high temperatures,
higher doping levels, or for large values of $t'$, the charged
holon scattering would give rise to the temperature linear
resistivity.

In summary, we have discussed the effect of the additional second
neighbor hopping $t'$ on the charge dynamics of the $t$-$J$ model
in the underdoped regime within the fermion-spin theory. Our
results indicate that the conductivity spectrum of the $t$-$t'$-$J$
model shows the low-energy peak and unusual midinfrared band, while
the resistivity exhibits a nearly temperature linear dependence
with deviation at low temperature in the underdoped regime.
Although the qualitative feature of the charge dynamics in the
$t$-$t'$-$J$ model is the same as in the case of the $t$-$J$ model,
the additional second neighbor hopping $t'$ leads to a clear shift
of the position of the midinfrared band to the higher energies in
the conductivity spectrum, and suppress the range of the deviation
from the temperature linear dependence in the resistivity.

Finally, we also note that some physical properties of the
$t$-$t'$-$J$ model have been discussed by Martins {\it et al}.
\cite{n20}, they found that there is a tendency of holes to
generate nontrivial spin environments, this effect leads to
decouple charge from spin. Moreover, they \cite{n20} have shown
that introducing $t'$ in the $t$-$J$ model is equivalent to
effectively renormalizing $J$ (decreasing the magnetic energy).
Therefore, within the charge-spin separation of the $t$-$t'$-$J$
model, the scattering of spinons dominates the spin dynamics
\cite{n13}, while the scattering of holons dominates the charge
dynamics, the two rates observed in the experiments are attributed
to the scattering of two distinct excitations, spinons and holons.

\acknowledgments
This work was supported by the National Natural Science Foundation
of China under Grant No. 10074007, 90103024, and 10125415, and the
Grant from Ministry of Education of China.

\begin{figure}
\caption{The conductivity of the $t$-$t'$-$J$ model with doping
concentration (a) $\delta=0.06$ and (b) $\delta=0.12$ for
parameters $t/J=2.5$, $t'/J=0.1$ (dashed line), $t'/J=0.3$ (dotted
line), and $t'/J=0.5$ (dash-dotted line) in temperature $T=0$.
The solid line is the corresponding result of the $t$-$J$ model.}
\end{figure}

\begin{figure}
\caption{The resistivity at doping concentration $\delta=0.06$
for parameters $t/J=2.5$, (a) $t'=0$ and (b) $t'/J=0.1$.}
\end{figure}

\begin{figure}
\caption{The holon density of states for parameters $t/J=2.5$
and $t'=0$ at doping $\delta=0.06$. The dashed line is the result
at the mean-field level.}
\end{figure}

\begin{figure}
\caption{The normal-state phase diagram $T^{*}\sim \delta$ for
parameters $t/J=2.5$ and $t'=0$. $T^{*}$ marks the development of
the holon pseudogap in the holon density of states}
\end{figure}

\begin{figure}
\caption{The normal-state phase diagram $T^{*}\sim t'$ at doping
$\delta=0.06$ for parameter $t/J=2.5$.}
\end{figure}

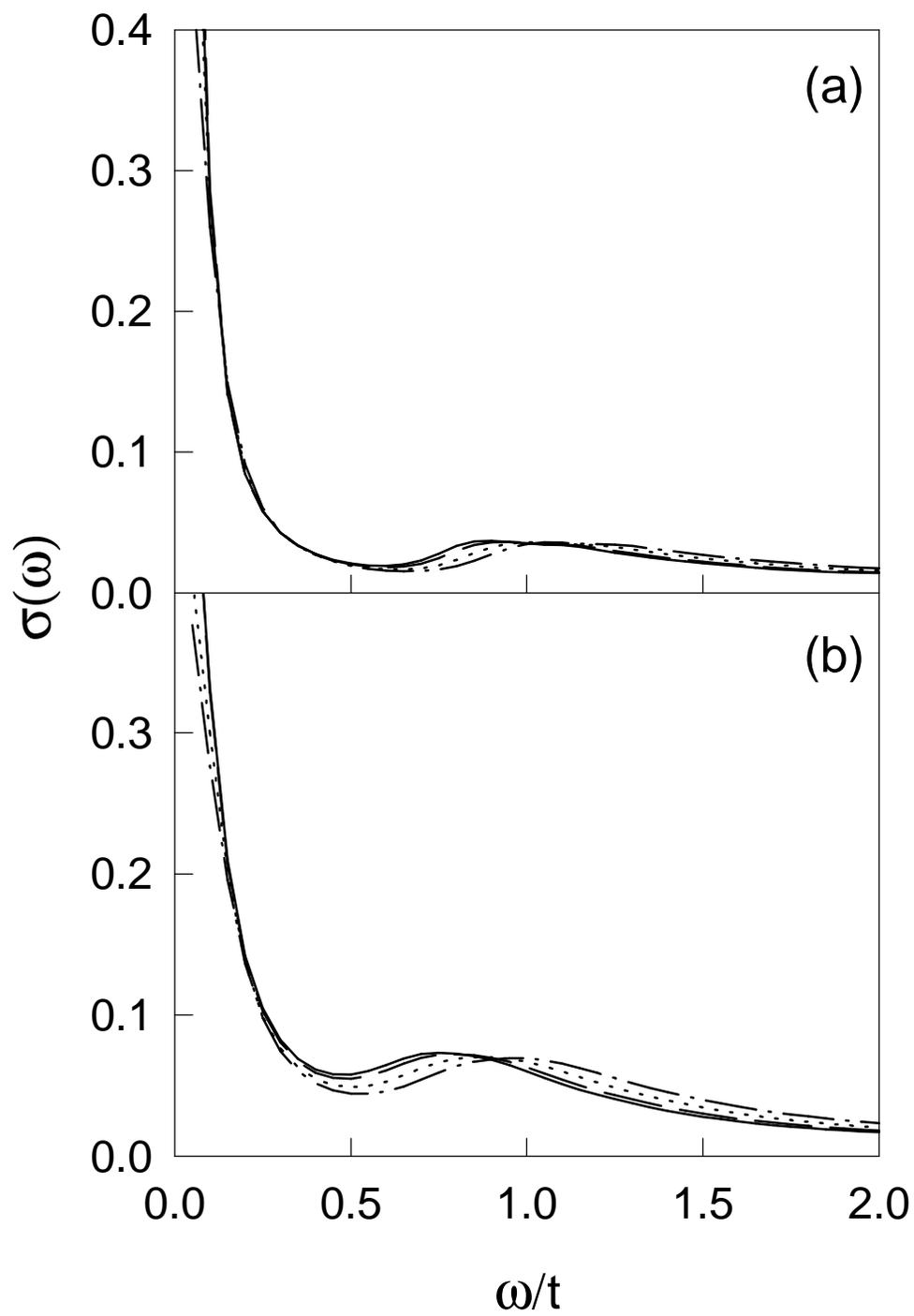
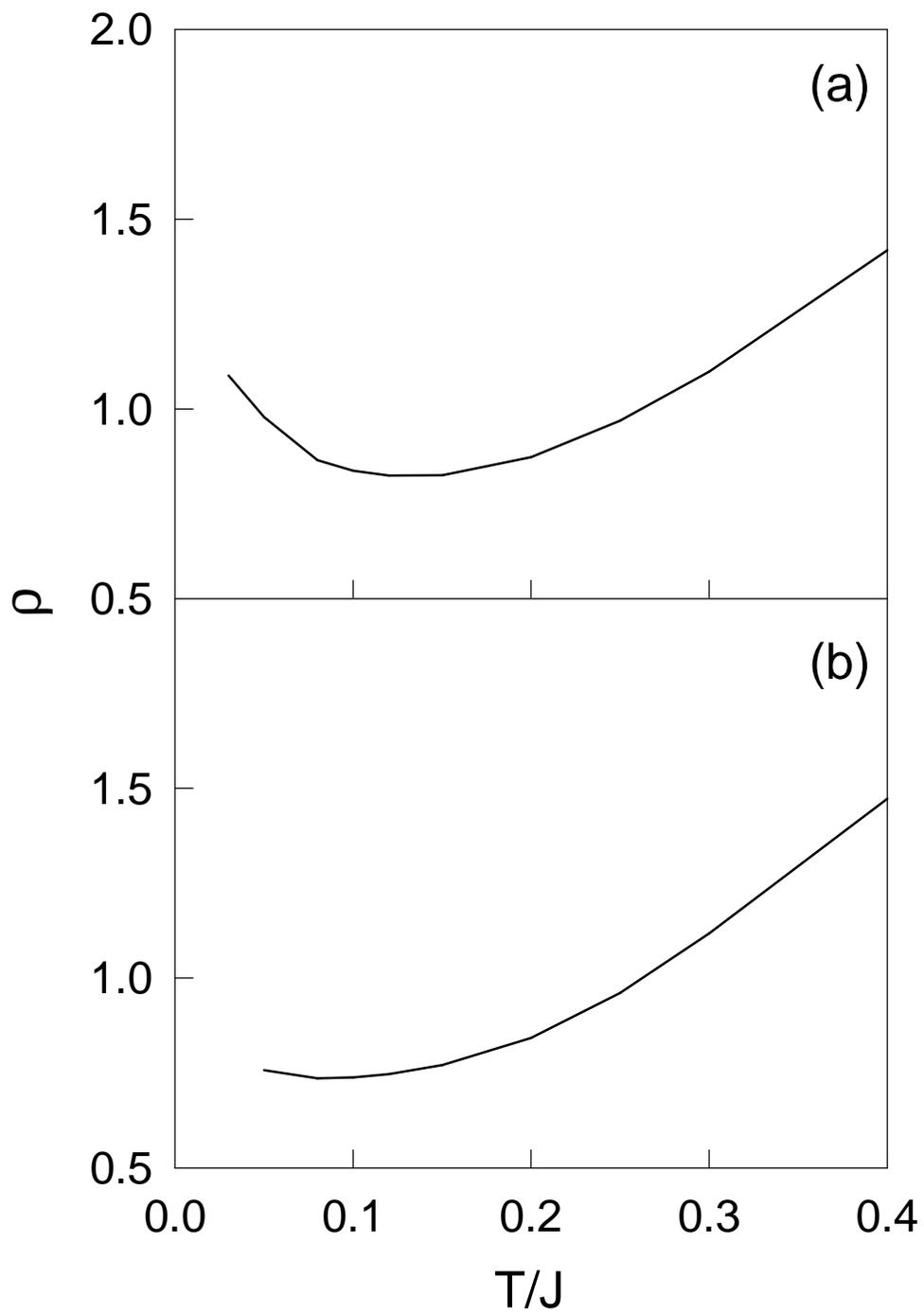
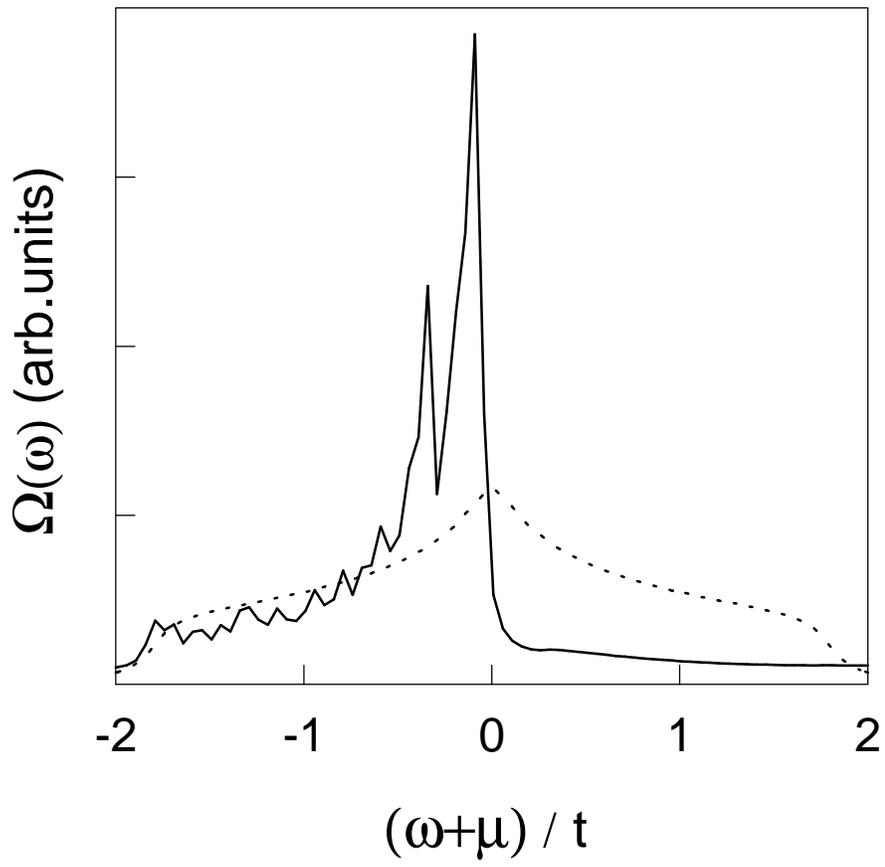
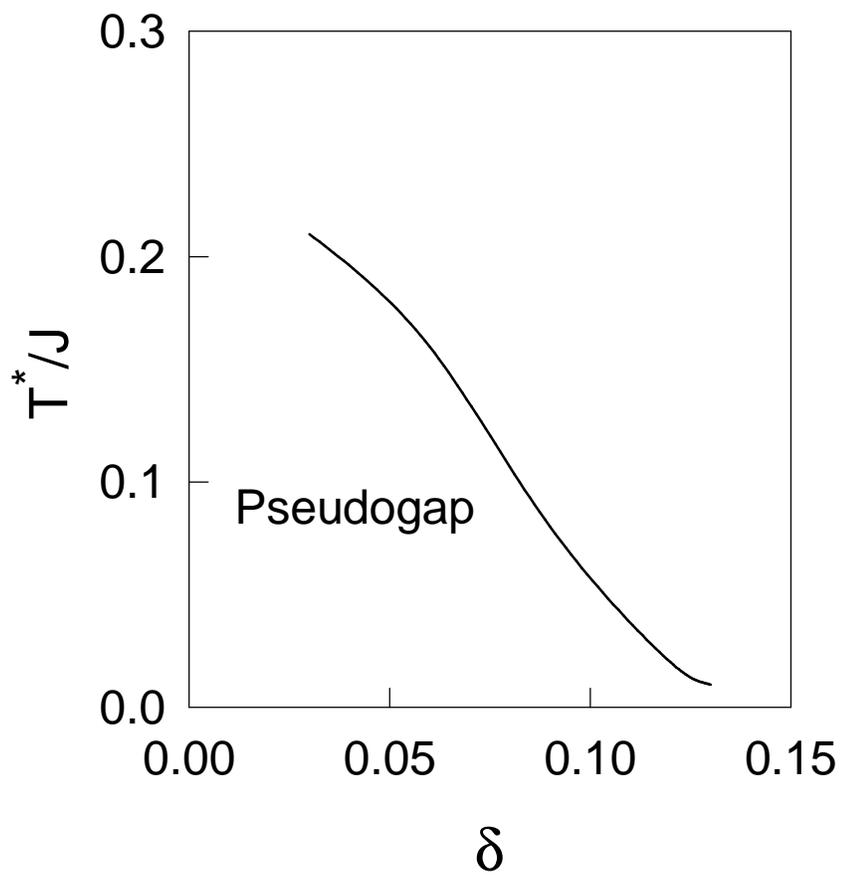
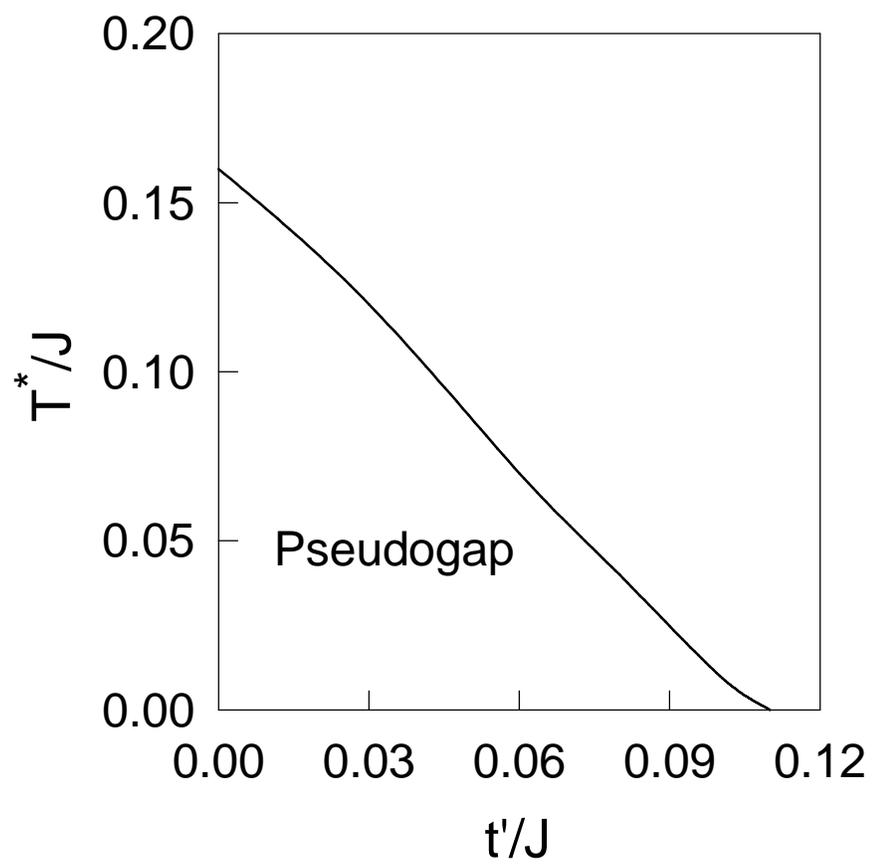
\end{document}